\documentclass[10pt,conference]{IEEEtran}
\IEEEoverridecommandlockouts
\usepackage{cite}
\usepackage{amsmath,amssymb,amsfonts}
\usepackage{algorithmic}
\usepackage[dvipdfmx]{graphicx}
\usepackage{subcaption}
\usepackage{textcomp}
\usepackage{xcolor}
\def\BibTeX{{\rm B\kern-.05em{\sc i\kern-.025em b}\kern-.08em
    T\kern-.1667em\lower.7ex\hbox{E}\kern-.125emX}}

\usepackage{eso-pic}
\newcommand\AtPageUpperMyright[1]{\AtPageUpperLeft{
 \put(\LenToUnit{0.02\paperwidth},\LenToUnit{-1cm}){
     \parbox{1.07\textwidth}{\raggedleft\fontsize{7}{11}\selectfont #1}}
 }}
\newcommand{\conf}[1]{
\AddToShipoutPictureBG*{
\AtPageUpperMyright{#1}
}
}

\begin{document}
\title{DeepSIP: A System for Predicting Service Impact of Network Failure by Temporal Multimodal CNN}
\author{\IEEEauthorblockN{Yoichi Matsuo\IEEEauthorrefmark{1}, Tatsuaki Kimura\IEEEauthorrefmark{2}, Ken Nishimatsu\IEEEauthorrefmark{1}}
\IEEEauthorblockA{\IEEEauthorrefmark{1} NTT Network Technology Laboratories, NTT Corporation\\
Email:{\{yoichi.matsuo.ex, ken.nishimatsu.hd}\}@hco.ntt.co.jp}
\IEEEauthorblockA{\IEEEauthorrefmark{2} Department of Information and Communications Technology, Graduate School of Engineering, Osaka University\\
Email:kimura@comm.eng.osaka-u.ac.jp}
}
\conf{ \copyright 2020 IEEE.  Personal use of this material is permitted.  Permission from IEEE must be obtained for all other uses, in any current or future media, including reprinting/republishing this material for advertising or promotional purposes, creating new collective works, for resale or redistribution to servers or lists, or reuse of any copyrighted component of this work in other works.}

\maketitle

\begin{abstract}
When a failure occurs in a network,
network operators need to recognize service impact, since service impact is essential information for handling failures.
%However, there have been a few studies on service impact comparing with anomaly detection or estimation of failed equipment.
%In this paper, we propose {\bf Deep} learning based {\bf S}ervice {\bf I}mpact {\bf P}rediction (DeepSIP),
%a system to predict the service impact of network failure in a network element using a temporal multimodal convolutional neural network (CNN).
%More precisely, DeepSIP predicts the time to recovery from the failure and the loss of traffic volume due to the failure in a network on the basis of information from syslog messages and traffic volume.
In this paper, we propose {\bf Deep} learning based {\bf S}ervice {\bf I}mpact {\bf P}rediction (DeepSIP),
a system to predict the time to recovery from the failure and the loss of traffic volume due to the failure in a network element using a temporal multimodal convolutional neural network (CNN).
Since the time to recovery is useful information for a service level agreement (SLA) and the loss of traffic volume is directly related to the severity of the failures,
we regard these as the service impact.
%the time to recovery and the loss of traffic volume as the service impact.
The service impact is challenging to predict,
since a network element does not explicitly contain any information about the service impact.
%since it depends on types of network failures and traffic volume.
%Moreover, a network element does not explicitly contain any information about the service impact.
%However, syslog messages contain the hidden information of what types of failures occurred in the network element.
%Since the time to recovery and the loss of traffic highly depends on the types of failures and traffic volume, we use syslog messages and past traffic volume to predict the service impact.
Thus, we aim to predict the service impact from syslog messages and traffic volume by extracting hidden information about failures. 
%s also challenging to be analyzed since these data are multimodal and strongly correlated, and have temporal dependencies.
To extract useful features for prediction from syslog messages and traffic volume which are multimodal and strongly correlated, and have temporal dependencies, we use temporal multimodal CNN.
We experimentally evaluated DeepSIP and DeepSIP reduced prediction error by approximately 50\% in comparison with other NN-based methods with a synthetic dataset.
\end{abstract}


\section{Introduction}
Service impact, such as time to recovery due to a failure in a network and loss of traffic volume caused by a failure,
needs to be predicted by network operators.
%When a failure occurs in the network elements such as switches, routers, and servers,
%it is insufficient for network operators to only detect anomalies or find the failed element.
%For instance, network operators immediately have to inform customers when a network failure will be recovered.
This is because the available time of network service is guaranteed as a service level agreement (SLA) between network providers and customers.
%Additionally in large-scale networks, many failures occur in the network simultaneously.
Additionally, if network operators can recognize the service impact before recovering the network, they can prioritize the order of recovery.
For example, if a failure has no or little impact on a service, network operators do not have to conduct repairs immediately.
Therefore, the prioritization will contribute to reducing operating expenses.

Even if a network failure is caused by a single network element failing,
the service impact is challenging to predict for three reasons.
First, a network element does not explicitly contain any information about service impact due to a failure.
Second, the temporal characteristics of service impact are highly diverse depending on what type of network failure occurs.
For instance, a failed module of routers generates more service impact than link flapping since it takes more time to replace a module than to recover from link flapping by quickly rebooting.
Third, even if the same type of network failure occurs, the service impact is not always the same.
This is because the service impact also depends on both traffic volume at the failure and past traffic volume.
%, which represents the number of customers who connect to the network.

\begin{figure}
\centering
\includegraphics[width=0.45\textwidth]{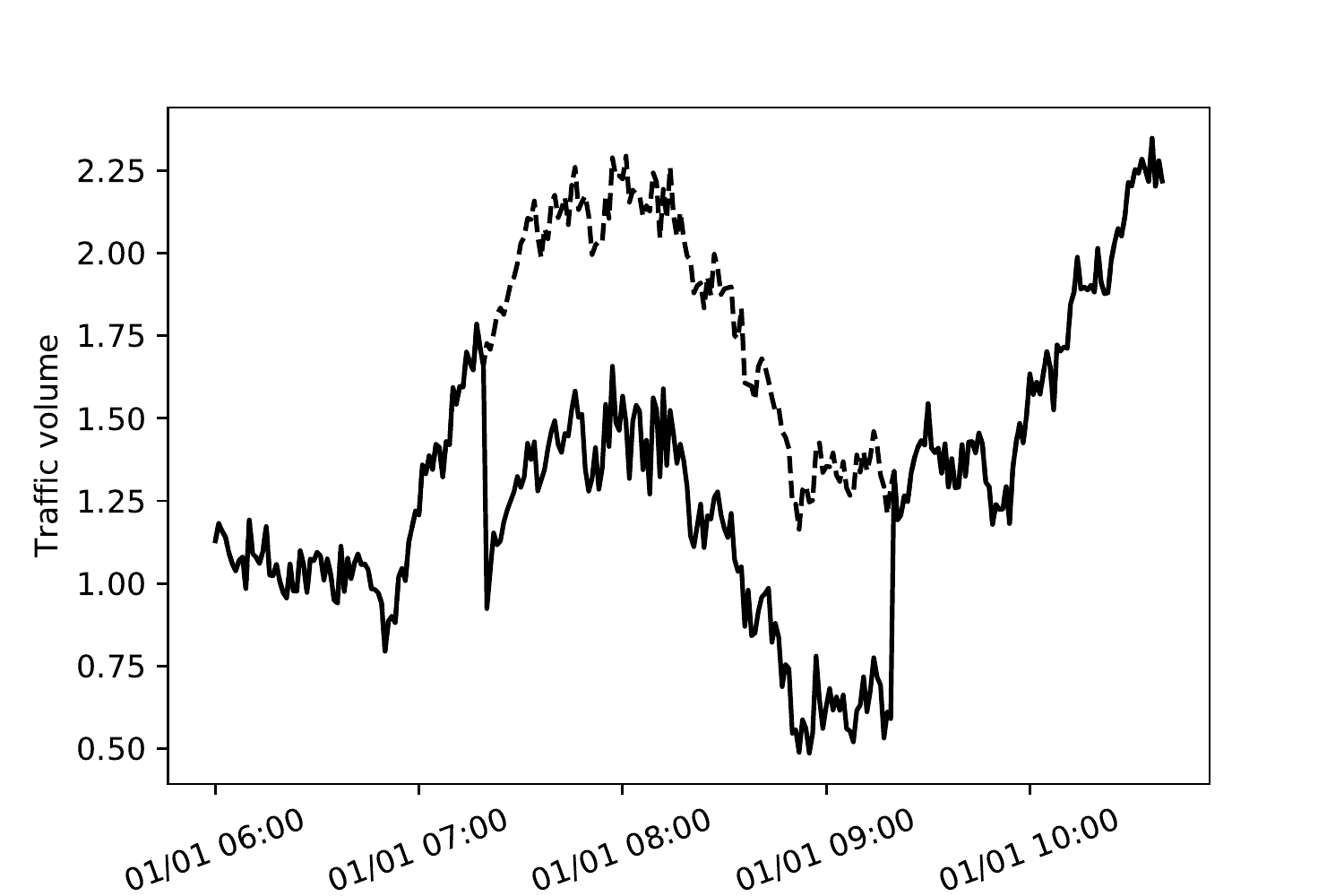}
\caption{Example of a service impact}
\label{fig:traffic_loss}
\end{figure}

We propose DeepSIP\footnote{DeepSIP means {\bf Deep} learning based {\bf S}ervice {\bf I}mpact {\bf P}rediction system}, a system to automatically predict the time to recovery from the failure and the loss of traffic volume due to the failure in a network element from syslog messages and traffic volume using temporal multimodal convolutional neural network (CNN).
%More precisely, DeepSIP predicts the time to recovery from the failure and the loss of traffic volume due to the failure in a network.
Since the loss of traffic volume is directly related to the severity of the failures, and represents the impact of the failures on customers, we regard the loss of traffic volume and time to recovery as types of service impact.
Figure~\ref{fig:traffic_loss} is an example of what DeepSIP predicts.
In this figure, a failure occurred at around 7:20 and recovery finished at around 9:20.
The solid line is actual traffic volume and the dashed line is traffic volume if the failure had not occurred.
DeepSIP predicts the time to recovery which is 120 min, and the loss of traffic volume, which is the area between the solid and dashed lines in this example.
%Since time to recovery from a failure and the loss of traffic volume represent the severity of the failures or impacts of a failure, they can be regard as service impact.

The time to recovery and the loss of traffic volume are challenging to predict since the network element do not explicitly contain information about service impact.
However, we consider that syslog messages and traffic volume may contain such information.
Since syslog messages describe what occurs in the network element, they contain the hidden information of what types of failures occur in network elements.
Since the time to recovery depends on recovery procedures, the types of failures are useful information.
For example, link flapping can be recovered by rebooting the interface, whereas a module fault can be recovered by replacing the module.
%These recovery procedures are described in trouble tickets so that every operator can recover a failed network element with the same procedure.
%Since the time to recovery depends on recovery procedures, it can be predicted.
The loss of traffic volume depends on how much traffic volume is reduced due to the failure and the time to recovery.
Thus, we argue that the time to recovery and the loss of traffic can be predicted by information of a failure type inferred from syslog messages and traffic volume at the time before a failure occurs.

To extract useful information for prediction from syslog messages and traffic volume are also challenging.
First, network data are multimodal, which means there are continuous values such as traffic volume and text data such as syslog messages.
We use a CNN to separately extract useful features using preprocessed vectors and scalar values generated from syslog messages and traffic volumes separately.
Second, syslog messages and traffic volume are strongly correlated.
%For instance, when syslog messages that represent a link-flap event appear, traffic volume will decrease.
Thus, we use another CNN to extract the correlation between syslog messages and traffic volume.
Finally, syslog messages and traffic volume have long and short temporal dependencies.
%These dependencies are difficult to explicitly put into a model for prediction since traffic volume contains many types of temporal dependencies.
Therefore, we again use a CNN for time-series data that can extract these dependencies automatically.
We evaluate the effectiveness and the accuracy of DeepSIP by comparing it with other NN-based methods.

The rest of this paper is organized as follows.
We describe related works in Section~\ref{sec:related_work}.
In Section~\ref{sec:proposal_method}, we explain DeepSIP, a system to predict service impact.
We evaluate DeepSIP in Section~\ref{sec:evaluation}.
Finally we conclude this paper in Section~\ref{sec:conclusino}.

\section{Related Work}
\label{sec:related_work}
\subsection{Network Operation}
Many studies about network operations such as anomaly detection, root cause analysis and recovery from the failures have been developed~\cite{Mahimkar2011a,Yoichi,deeplog}.
PRISM~\cite{Mahimkar2011a} is a system for detecting performance change of network elements due to maintenance by singular value decomposition.
%Deep learning based methods have also been proposed in a survey~\cite{ahmed2016survey} for anomaly detection.
%Aygun and Yavuz~\cite{aygun2017network}, proposed an autoencoder based model to detect security incidents in network elements.
%In the survey~\cite{ahmed2016survey}, many anomaly detection methods using deep learning methods are proposed.
%For root cause analysis and fault localization, methods based on a stochastic model~\cite{Yoichi, Bennacer2015} have been proposed.
%Shrink~\cite{Kandula2005} and Gestalt~\cite{Mysore2014} localize fault element in the IP network using graphical modeling.
%Shrink~\cite{Kandula2005} localizes a fault element in a IP network by graphical modeling.
%Root cause analysis systems based on a combination of case based reasoning and a stochastic model have also been proposed~\cite{Yan2012a, Bennacer2015}.
%These methods and systems find a root cause using the causal relationships between a failed network element and observation data, such as syslog messages and traffic volume.
%DeepLog~\cite{deeplog} detects anomalies and estimates the root causes of anomalies by a long short-term memory (LSTM)~\cite{lstm} based model using syslog messages.
%For recovery task, analyzing and Visualizing unstructured trouble tickets method~\cite{watanabe2018} to extract workflows is proposed for recovery recommendation.
%Since extracted workflows indicate recovery procedures from failures, this method makes recovery task stable.
However, only a study~\cite{Yang} have focused on service impact despite its importance.
Yang et al.~\cite{Yang} considered the number of affected user devices as service impact and modeled how many user devices are affected in a cellular environment when a base station fail.
Unlike their analysis~\cite{Yang}, our system takes into account the difference in types of failures to predict the time to recovery and the loss of traffic volume.

\subsection{Deep Learning Algorithm for Time-series Data}
Deep learning based regression methods have been proposed in recent years.
Representative methods for time-series data, are a long short-term memory (LSTM)~\cite{lstm}.
%LSTM was developed to overcome the vanishing or exploding gradient problem that made temporal dependencies impossible to extract.
However, it has been reported that LSTM cannot extract long-time temporal dependencies in the data~\cite{bradbury2016quasi}.
%To deal with time-series data with long time dependencies, QRNN and WaveNet are proposed.
Quasi-recurrent NN (QRNN)~\cite{bradbury2016quasi} proposed to deal with long time dependencies.
Applying a CNN to time-series data enables us to directly input whole time-series data.
% while we input each time-series data to LSTM.
%It uses previous time-series data for predicting data at next time.
%The filters of a CNN learn temporal dependencies between time-series data.
%WaveNet~\cite{wavenet} is also CNN-based method for analyzing audio data.
%To learn very long time dependencies.
Traffic volume prediction method using deep learning for time-series data have also been developed~\cite{zhang2019deep}.
Wang et al.~\cite{wang2017spatiotemporal} proposed a combination of an autoencoder and LSTM model to predict traffic volumes of each link in the network.
%A graph neural network model has been proposed to learn spatial-temporal dependencies of traffic volumes in a mobile network~\cite{wang2018spatio}.
%By using previous data directly, these methods can learn long time dependencies.

However, the above methods focus only on unimodal such as audio data and traffic data.
Regarding multimodal data, DeepSense~\cite{deepsense} was proposed for classification or regression tasks using time-series sensor data.
It combines a CNN and gated recurrent unit (GRU), which is an LSTM-based method.
%However, it has been reported that RNN based methods can not extract long time dependencies~\cite{bradbury2016quasi}.
However, since LSTM cannot extract long-time temporal dependencies, we propose applying QRNN instead, to extract long-time temporal dependencies from the multimodal time-series data.

\section{DeepSIP}
\label{sec:proposal_method}
In this section, we explain our system, DeepSIP.
In short, when a network element fails, DeepSIP predicts the service impact of the failure, i.e., the time to recovery from the failure and the loss of traffic volume due to the failure, using past syslog messages and traffic volume of the network element.
The loss of traffic volume is directly related to the severity of the failures and represents the impacts of the failures on customers using the network at that time.
%represents the impacts of the failures on customers, we regard the loss of traffic volume and time to recovery as types of service impact.
%More precisely, by combining the time to recovery, we can calculate the average loss of traffic volume during the failures.
Moreover, the approximate number of affected customers can be estimated using the average loss of traffic volume which is calculated by the time to recovery, and average sending or receiving traffic volume per customer.
Therefore, DeepSIP predicts the time to recovery and the loss of traffic volumes.

%To overcome the problem of LSTM, we apply QRNN to extract long-time temporal dependencies from the multimodal time-series data.
Since syslog messages and traffic volume are time-series data with different features,
which means that there are continuous values such as traffic volumes and text data such as syslog messages,
DeepSIP uses the CNN for multimodal data. 
Moreover, we propose applying QRNN instead of LSTM, to extract long-time temporal dependencies from the multimodal time-series data.
%syslog messages do not explicitly include information about service impact when a failure occurred.
%However, syslog messages contain the types of failures that occurred in a network element since they include various information about failures, such as hardware failures messages or link-down messages.
%We argue that the time to recovery from the failure can be predicted from the type of failures.
%The loss of traffic volume can be predicted by combining information on how much traffic volume there would be if the failure had not occurred, which is estimated by past traffic volume, and the type of failures.
%Therefore, DeepSIP predicts service impact by extracting information about the type of failure and traffic volume from syslog messages and past traffic volumes.

In the following subsections, we first define the considered problem.
Then, we give an overview of DeepSIP.
Finally, we give details of the temporal multimodal CNN in DeepSIP.
Notations used in this paper are summarized in Table~\ref{tab:notation}.
\begin{table}
\centering
\caption{Definition of symbols}
\label{tab:notation}
\begin{tabular}{c|c}
\hline
Symbol & Definition \\
\hline
\hline
$t$ & Time slot \\
\hline
$T$ & Time slot in which a failure occurred\\
\hline
$\mathsf{TTR}$ & Time slot in which recovery finished \\
\hline
$V_{\mathsf{TTR}}$ & Loss of traffic volume between $T$ and $\mathsf{TTR}$ \\
\hline
$\mathsf{TTR}^{\prime}, \hat{V}_{\mathsf{TTR}^{\prime}}$ & Actual data in the past \\
\hline
$\{ {\cdot}_t\}^T_{t=1}$ & Set of time-series data \\
\hline
$\boldsymbol{x}_t$ &
\begin{tabular}{c}
Vector at time $t$ \\ that represents preprocessed syslog messages
\end{tabular}
 \\
\hline
$\{\boldsymbol{x}_t\}^T_{t=1}$ & Time-series data of $\boldsymbol{x}_t$ \\
\hline
$y_t$ & Scalar value at time $t$ that represents traffic volume \\
\hline
$\mathcal{D}$ &
\begin{tabular}{c}
training dataset, \\
$\{(\{\boldsymbol{x}_t\}^T_{t=1}, \{ y_t \}^T_{t=1}, \mathsf{TTR}^{\prime}, \hat{V}_{T \rightarrow \mathsf{TTR}^{\prime}})_d \}^{|\mathcal{D}|}_{d=1}$
\end{tabular}
 \\
\hline
$\{ y_t \}^T_{t=1}$  & Time-series data of $y_t$ \\
\hline
$w_{i}$ & Weight of CNN in layer $i$ \\
\hline
$z^{(i)}$ & Output of CNN from layer $i$ \\
\hline
\end{tabular}
\end{table}

\subsection{Service Impact Prediction Problem}
We explain the problem formulation of predicting service impact.
Let $T$ be the time slot in which a failure occurred and $\mathsf{TTR}$ be the time slot in which recovery finished,
where time $t$ is slotted as $ t=1, 2, \ldots, T, \ldots, \mathsf{TTR}$.
Let $V_{\mathsf{TTR}}$ be the scalar value that represents loss of traffic volume between $T$ and $\mathsf{TTR}$ due to a failure.
%For simplification, we denote $V_{T \rightarrow \mathsf{TTR}}$ as $V_{\mathsf{TTR}}$.
We denote a set of time-series data as $\{ {\cdot}_t\}^T_{t=1}$.
Let $\boldsymbol{x}_t$ be a vector at time $t$ and $\{\boldsymbol{x}_t\}^T_{t=1}$ be time-series data of $\boldsymbol{x}_t$.
Similarly, let $y_t$ be a scalar value at time $t$ and $\{ y_t \}^T_{t=1}$ be a time-series data of $y_t$.
These vectors and scalar values are generated from syslog messages and traffic volume respectively, and input into the temporal multimodal CNN.
We explain how to generate these vectors and scalar values in Section~\ref{subsec:architecture_deepsip}.
The goal of DeepSIP is to predict the $\mathsf{TTR}$ and $V_{\mathsf{TTR}}$ from the collected syslog messages and traffic volume in the past and at the time when the failure occurs.
Even though DeepSIP uses traffic volume at $T$ as input, the loss of traffic volume $V_{\mathsf{TTR}}$ is difficult to predict
since it depends on the type of failure.
For instance, we cannot distinguish the difference in traffic changes in a network element between spike and packet drops when a failure has just occurred.
DeepSIP predicts them using syslog messages that contain the types of failures in network element.

\subsection{System Overview}
\begin{figure}
\centering
\includegraphics[width=0.4\textwidth]{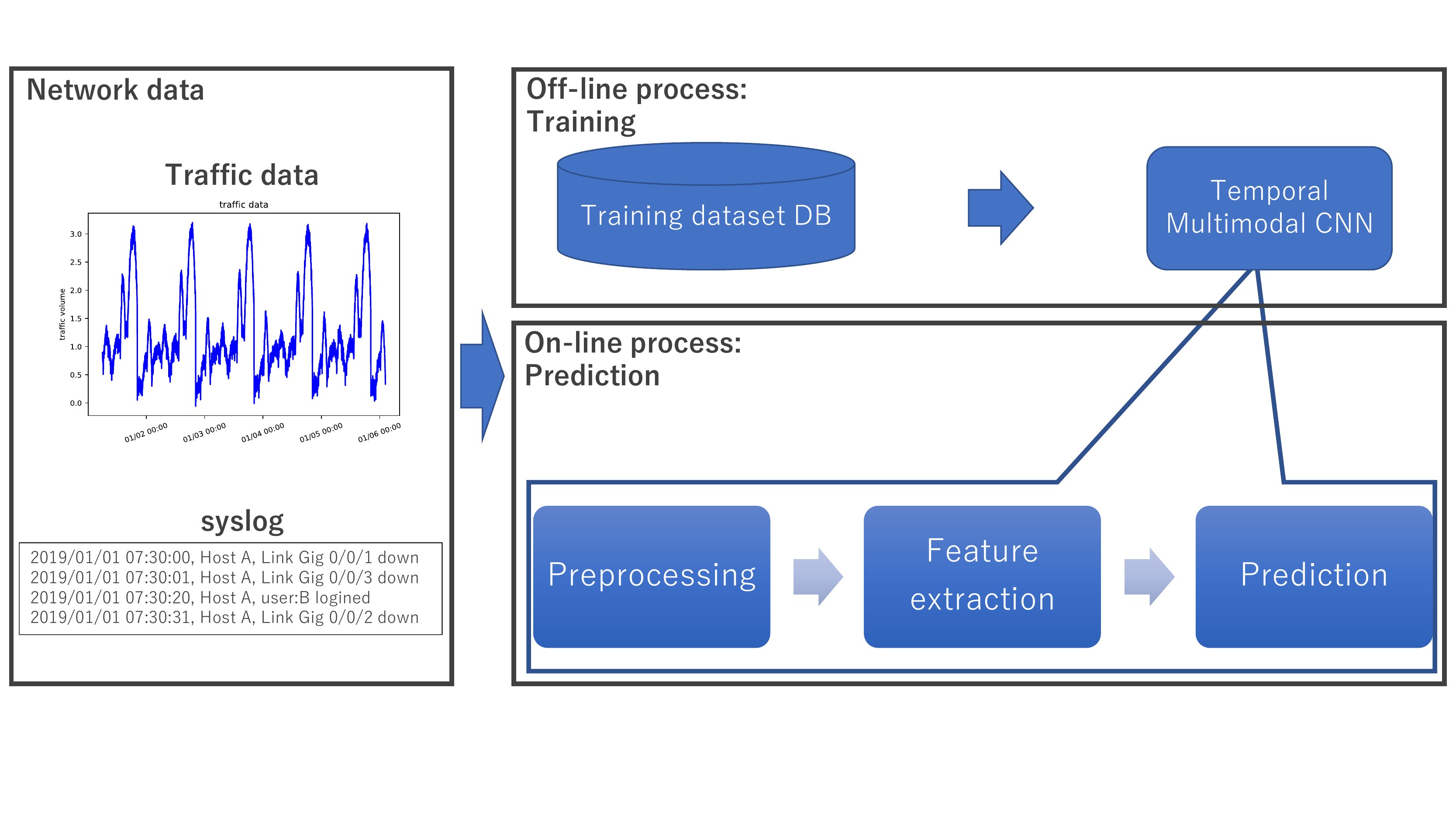}
\caption{System Overview of DeepSIP}
\label{fig:system_overview}
\end{figure}

We show the system overview of DeepSIP in Figure~\ref{fig:system_overview}.
DeepSIP consists of two phases: a training phase, which is an off-line process, and a predicting phase, which is an on-line process.
In the training phase, weights in DeepSIP are trained using training dataset $\mathcal{D}$.
Superscript ${ }^{\prime}$ denotes actual data from past failures in the training dataset.
Let ${\mathsf{TTR}^{\prime}}$ be a time to recovery at the past failure and ${V}^{\prime}_{{\mathsf{TTR}^{\prime}}}$ be a loss of traffic volume at the past failure.
Training data for DeepSIP consist of a tuple, $(\{\boldsymbol{x}_t\}^T_{t=1}, \{ y_t \}^T_{t=1}, {\mathsf{TTR}^{\prime}}, {V}^{\prime}_{\mathsf{TTR}^{\prime}})$,
and the $\mathcal{D}$ is a set of tuples, $\mathcal{D}=\{(\{\boldsymbol{x}_t\}^T_{t=1}, \{ y_t \}^T_{t=1}, {\mathsf{TTR}^{\prime}}, {V}^{\prime}_{T \rightarrow {\mathsf{TTR}^{\prime}}})_d \}^{|\mathcal{D}|}_{d=1}$.
Here $|\mathcal{D}|$ is the amount of data for training.
In the predicting phase, DeepSIP predicts $\mathsf{TTR}$ and $V_{\mathsf{TTR}}$ from the collected syslog messages and traffic volume.
%If there is not failure, it predicts zeros.
\begin{table*}[t!]
\centering
\caption{Examples of the templates of syslog messages}
  \begin{tabular}{l|c|l}
  \hline
  syslog messages & template ID & syslog template \\
  \hline\hline
  2019/01/01 07:30:00, Host A, Interface Gig 0/0/1 down & 1 & XXX, Host A, Interface Gig XXX down \\
  \hline
  2019/01/01 07:30:10, Host A, Interface Gig 0/0/1 state changed & 2 & XXX, Host A, Interface Gig XXX state changed \\
  \hline
  2019/01/01 07:30:20, Host A, user:B logged in & 3 & XXX, Host A, XXX logged in\\
  \hline
  2019/01/01 07:30:31, Host A, Interface Gig 0/0/2 down & 1 & XXX, Host A, Interface Gig XXX down \\
  \hline
  \end{tabular}
\vspace{5mm}
\label{tab:templation}
\end{table*}

\subsection{Temporal Multimodal CNN in DeepSIP}
\label{subsec:architecture_deepsip}
\begin{figure}
\centering
\includegraphics[width=0.4\textwidth]{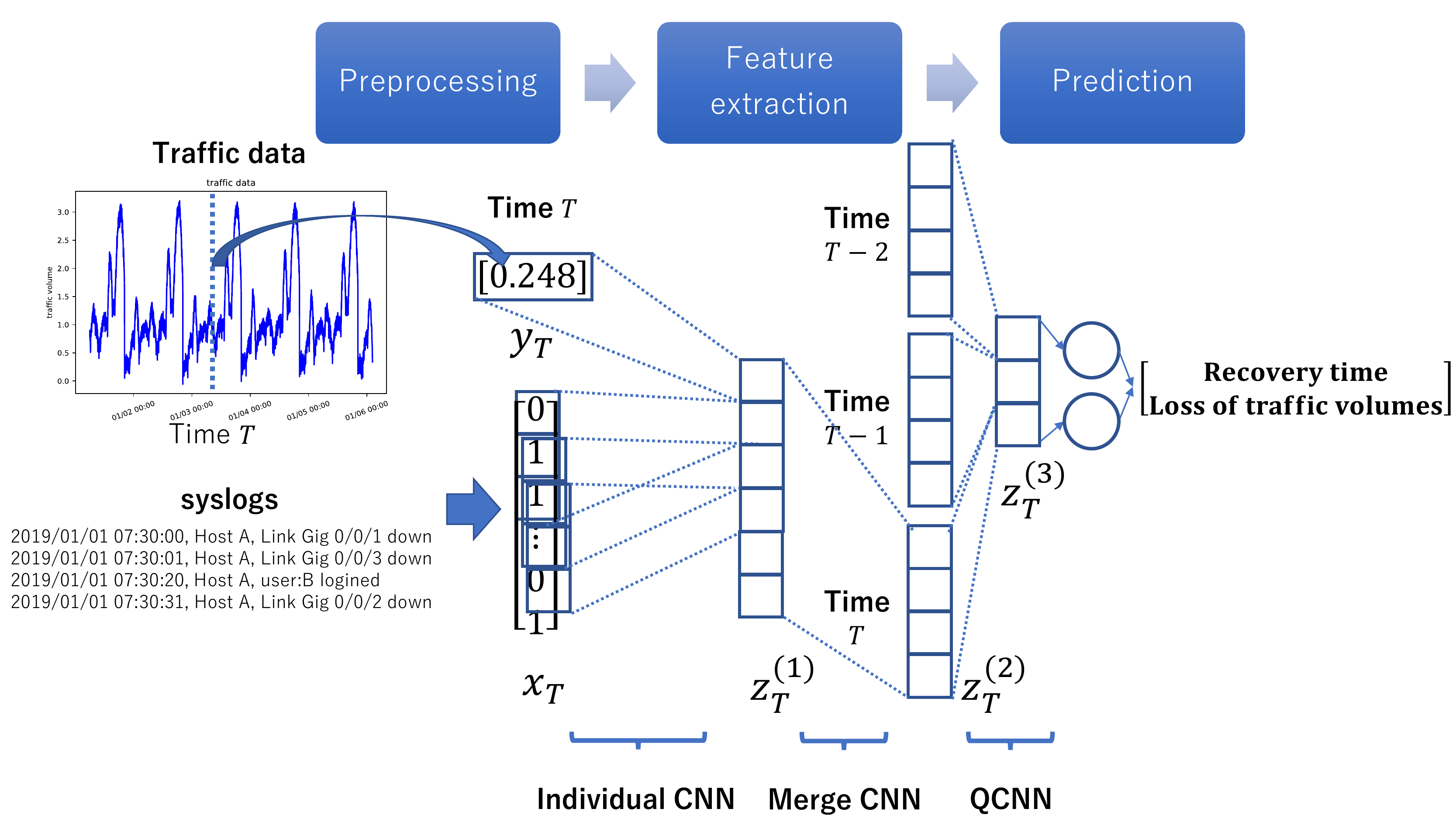}
\caption{Temporal multimodal CNN in DeepSIP}
\label{fig:architecture}
\end{figure}

Temporal multimodal CNN in DeepSIP consists of three parts: preprocessing, feature extraction, and prediction.
The temporal multimodal CNN in DeepSIP is illustrated in Figure~\ref{fig:architecture}. 
In the preprocessing part, DeepSIP generates preprocessed vectors $\{\boldsymbol{x}_t\}^T_{t=1}$ and scalar values $\{ y_t \}^T_{t=1}$ from syslog messages and traffic volume.
In the feature extraction part, DeepSIP extracts the features of $\boldsymbol{x}_t$ and $y_t$,
correlations between $\boldsymbol{x}_t$ and $y_t$,
and temporal dependencies of $\{\boldsymbol{x}_t\}^T_{t=1}, \{ y_t \}^T_{t=1}$ using CNNs.

\subsubsection{Preprocessing}
In the preprocessing part,
we first split syslog messages and traffic volume accordance with a time slot.
Then, to be able to handle syslog messages as numerical values, we use {\it template methods}~\cite{Kimura2015, deeplog}.
These methods are for automatically assigning a template ID to each message of the syslog.
If syslog messages have the same meaning except the numerical part, such as timestamps or IP addresses, they are given the same template ID even if the messages of syslog differ.
Examples of the templates are shown in Tab.~\ref{tab:templation}.

$M$ is the size of $\boldsymbol{x}_t$ and is the total number of template IDs.
Each element $x_{t,m}$ of $\boldsymbol{x}_t$ represents the number of occurrences of the $m$-th template ID within a time slot, $\tau$.
Similarly, $y_t$ represents the traffic volume at time $t$.
$\{\boldsymbol{x}_t\}^T_{t=1}$ and $\{ y_t \}^T_{t=1}$ are the preprocessed vectors and scalar values, respectively, which are input for the temporal multimodal CNN.

\subsubsection{Feature Extraction}
In the feature extraction part, DeepSIP extracts features of syslog messages and traffic volume separately using CNNs.
DeepSIP also extracts correlations between syslog messages and traffic volume, and time dependencies using CNNs.
First, DeepSIP extracts internal features from multimodal data which are a preprocessed vector $\boldsymbol{x}_t$ and scalar value $y_t$ at time $t$ separately.
%This is because the same values of $\boldsymbol{x}_t$ and $y_t$ have different meanings.
%Therefore we should use extracted features values for extracting patterns instead of $\boldsymbol{x}_t$ and $y_t$.
We use 1d CNN, a CNN for vectors, for this extraction.
%Note that we do not explain the detail of CNN in this paper.
%We use no padding, and input and output channels are set to 1.
The output vector $\boldsymbol{z}^{(1,\mathrm{syslog})}_t$ and its $j$-th element $z^{(1,\mathrm{syslog})}_{t,j}$ can be calculated by the following equations.
\begin{eqnarray}
\boldsymbol{z}^{(1,\mathrm{syslog})}_{t} = \mathrm{1dCNN}^{(\mathrm{syslog})}(\boldsymbol{x}_t), \\
z^{(1,\mathrm{syslog})}_{t,j} = \sum^{K}_{k=0} w^{(\mathrm{syslog})}_{k} \times x_{t,k+j},
\end{eqnarray}
where $K$ is kernel size i.e., how many elements CNN convolutes, and $w^{(\mathrm{syslog})}_{k}$ are the weights of the CNN that are trained.
The size of output vector $\boldsymbol{z}^{(1,\mathrm{syslog})}_{t}$ is $M - K$.
%In our system,
%For more detail of CNN implementation, please see~\cite{}.
Similar to $\boldsymbol{z}^{(1,\mathrm{syslog})}_{t}$, $\boldsymbol{z}^{(1,\mathrm{traffic})}_{t}$ is also calculated by 1d CNN.
\begin{equation}
\boldsymbol{z}^{(1,\mathrm{traffic})}_{t} = \mathrm{1dCNN}^{(\mathrm{traffic})}(y_t),
\end{equation}
where we set the kernel size as 1.
We denote this layer of CNN as an individual the CNN following DeepSense.

Second, DeepSIP extracts the correlation between syslog and traffic volumes at time $t$.
We arrange extracted features $\boldsymbol{z}^{(1,\mathrm{syslog})}_{t}$ and $\boldsymbol{z}^{(1,\mathrm{traffic})}_{t}$ into a vector $\boldsymbol{z}^{(1)}_t$ with size $M - K +1$.
We again use 1d CNN for this extraction.
\begin{eqnarray}
\boldsymbol{z}^{(2)}_t = \mathrm{1dCNN}(\boldsymbol{z}^{(1)}_t), \quad
z^{(2)}_{t,j} = \sum^{K}_{k=0} w^{(2)}_{k} \times z^{(1)}_{t,k+j},
\end{eqnarray}
where $w^{(2)}_{k}$ are the weights of the CNN.
We also denote this layer of the CNN as merge CNN following DeepSense.

Finally, DeepSIP extracts time dependencies of $\boldsymbol{z}^{(2)}_t$ between time series from 1 to $T$ using QRNN.
We arrange $\{\boldsymbol{z}^{(2)}_t\}^T_{t=1}$ in the column direction
to create a matrix $Z^{(2)} \in \mathbb{R}^{N \times T}$, where $N$ is the size of $\boldsymbol{z}^{(2)}_t$.
DeepSIP outputs $Z^{(3)}$ using QRNN.
\begin{eqnarray}
Z^{(3)} = \mathrm{QRNN}(\boldsymbol{z}^{(2)}_1, \ldots, \boldsymbol{z}^{2}_T ),\quad
z^{(3)}_{t,j} = \sum^{T}_{t=1} w^{(3)}_{t} \times \boldsymbol{z}^{(2)}_{t,t+j},
\end{eqnarray}
where $w^{(3)}_{t}$ are weights of QRNN.

\subsubsection{Prediction}
In the prediction part, DeepSIP predicts time to recovery $\mathsf{TTR}$ and the loss of traffic volume $V_{\mathsf{TTR}}$ using $Z^{(3)}$
which has temporal and multimodal features in the syslog messages and traffic volumes.
Predicting $\mathsf{TTR}$ and $V_{\mathsf{TTR}}$ is a regression task.
This is done by linear functions, $F_{TTR}$ and $F_{V}$, as in the following equations.
\begin{equation}
\mathsf{TTR} = F_{\mathsf{TTR}}(\boldsymbol{Z}^{(3)}), \quad
V_{\mathsf{TTR}} = F_{V}(\boldsymbol{Z}^{(3)}),
\end{equation}
When we train the model, we use the mean squared error as the loss function and update every weight of DeepSIP using SDG.
%Then DeepSIP can predict service impact after training phase.

Note that the architecture of DeepSIP can be used for general purposes by changing the function in the prediction part.
For classification, such as predicting the type of the failure, we use a softmax function to normalize the sum of the $\boldsymbol{Z}^{(3)}$ values to 1.
% so that $\boldsymbol{Z^{(3)}}$ can be seen as probability.
%In case of prediction of $\boldsymbol{x}_{t+1}, y_{t+1}$, we use a linear function as same as the regression.
By using this architecture,
it is possible to extract the features of dependencies in the time series,
those of each data and those between data for multimodal data.

\section{Evaluation}
\label{sec:evaluation}
In this section, we evaluate of DeepSIP by using synthetic data that imitate realistic traffic volume and syslog messages.
To extensively evaluate the accuracy of estimated service impacts by DeepSIP,
we comprehensively generated synthetic traffic volumes and syslog messages,
and injected degradation to these synthetic data to imitate the behavior of traffic volumes and syslog messages when failures occur.
For realistic failures and corresponding degradation of traffic volumes, we referred to the idea about degradation patterns considered by Mahimkar et al.~\cite{Mahimkar2011a}.
% which occurred in the network elements and prepared data when these failures occurred.
Using these data, we compared DeepSIP using the temporal multimodal CNN with base-line methods.

In what follows, first, we explain degradation patterns of traffic volumes when failures occur in the network.
Then, we explain how to generate synthetic traffic volumes and syslog messages, and inject degradation at the time a failure occurs.
Finally, we show experiment results and evaluations.

\subsection{Failures and Corresponding Degradation Patterns}
\label{subsec:degradation_patterns}
\begin{figure*}
\centering
\begin{tabular}{cccc}
\centering
\begin{minipage}[b]{0.24\linewidth}
\centering
\includegraphics[width=0.9\textwidth]{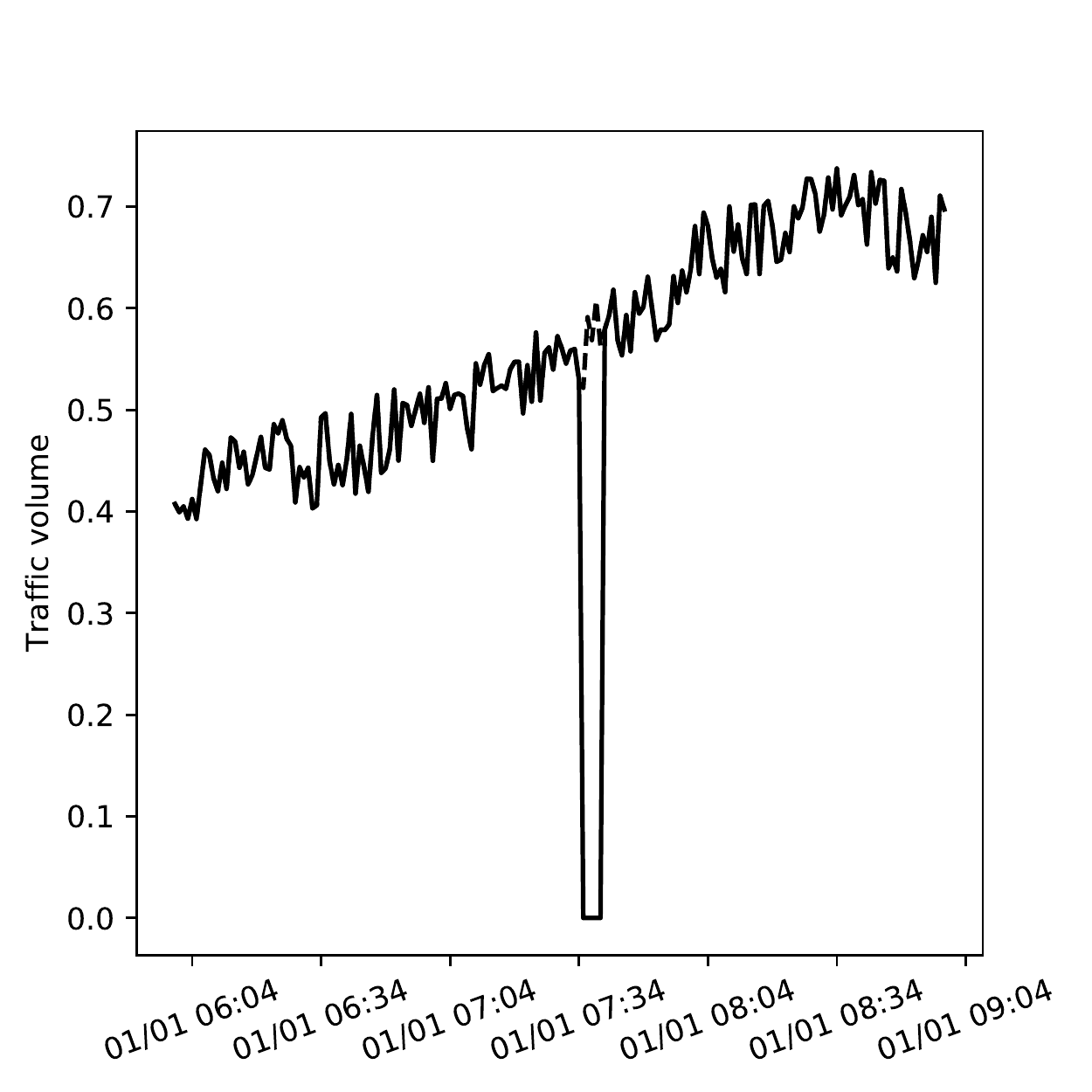}
\subcaption{Spike}
\label{fig:spike}
\end{minipage} &
\begin{minipage}[b]{0.24\linewidth}
\centering
\includegraphics[width=0.9\textwidth]{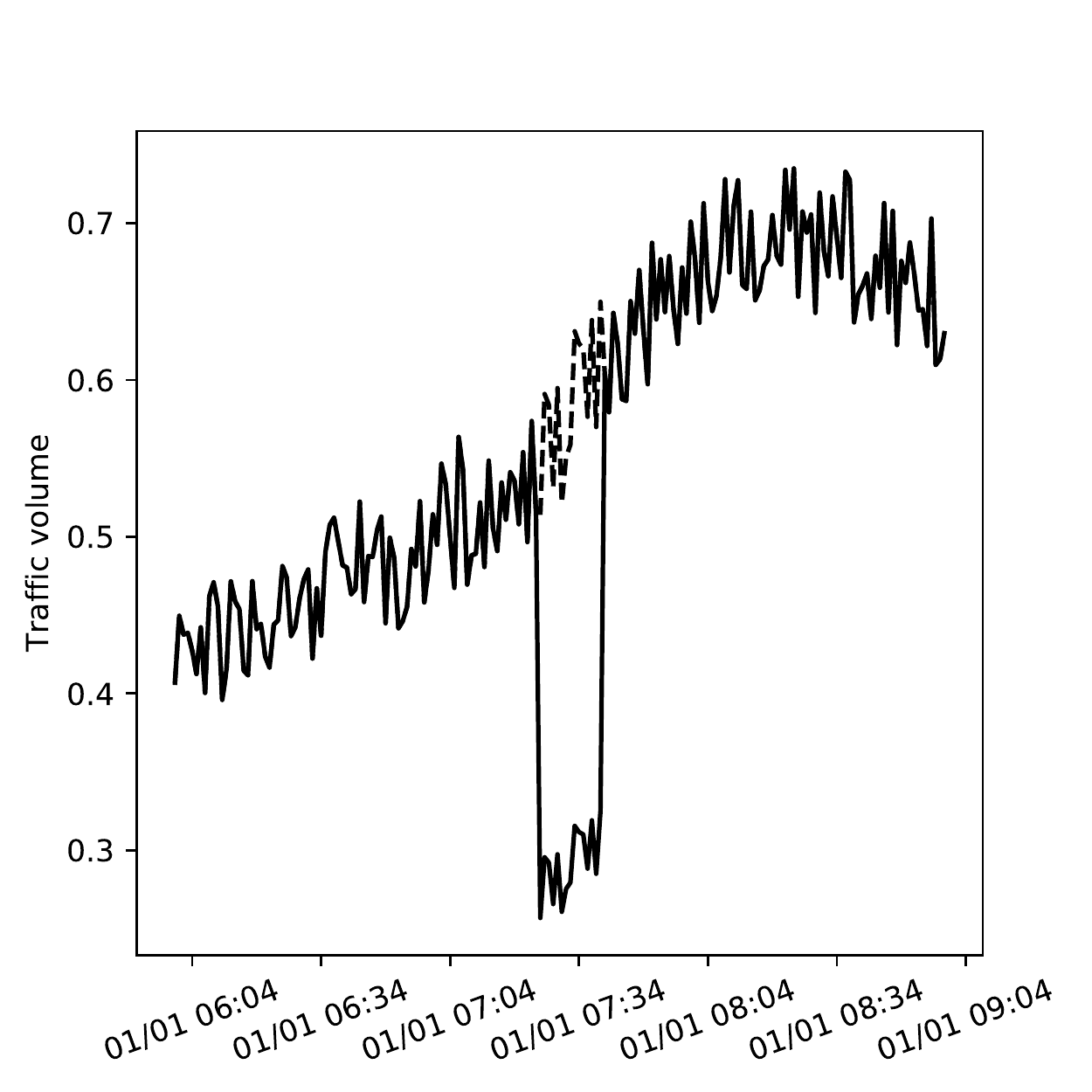}
\subcaption{Level shift}
\label{fig:levelshift}
\end{minipage} &
%\end{figure}
\begin{minipage}[b]{0.24\linewidth}
%\begin{figure}
\centering
\includegraphics[width=0.9\textwidth]{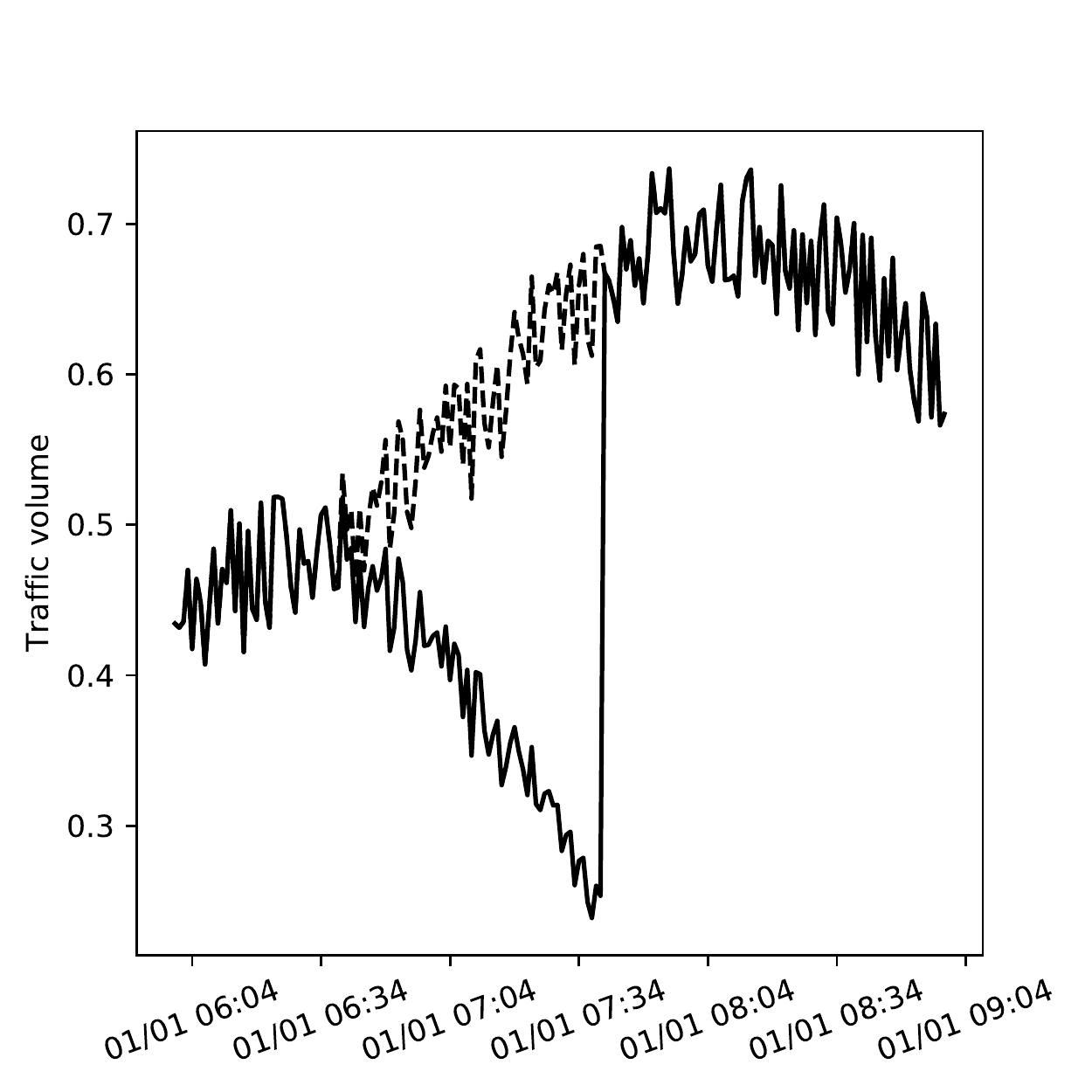}
\subcaption{Ramp down}
\label{fig:rampdown}
\end{minipage}
%\end{figure}
%\begin{figure}
\begin{minipage}[b]{0.24\linewidth}
\centering
\includegraphics[width=0.9\textwidth]{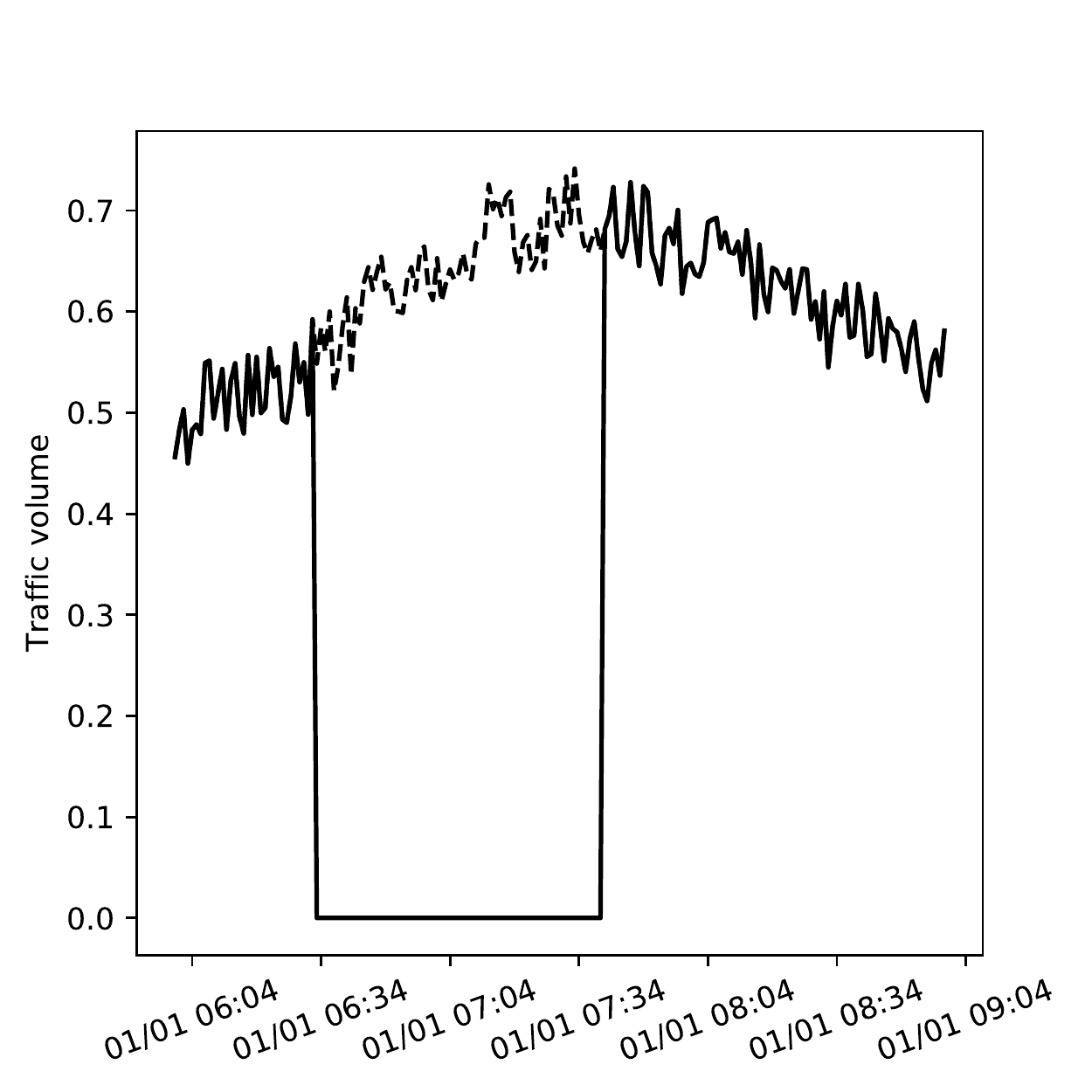}
\subcaption{Long down}
\label{fig:full}
\end{minipage}
\end{tabular}
\caption{Traffic pattern of failures}
\label{fig:traffic_pattern}
\end{figure*}

In this subsection, we explain traffic behaviors when failures occur.
For evaluation, Mahimkar et al.~\cite{Mahimkar2011a} introduced realistic failures and corresponding degradation patterns of traffic volume, which are {\it spike}, {\it level-shift} and {\it ramp down}.
% which are highly desirable to detect for network operations.
We show examples of each degradation pattern of traffic volumes in Figures~\ref{fig:traffic_pattern}.
The first pattern is spike in Figure~\ref{fig:traffic_pattern} (a), which completely reduces traffic volume.
An example of the failures that cause spike pattern is link flap.
This is because link flap stops a network element from connecting to other network elements,
but it is recovered quickly by rebooting the interface.
%Such traffic behavior occurs when a link flap occurs.
The second pattern is level shift in Figure~\ref{fig:traffic_pattern} (b), which reduces a certain amount of traffic volume during a certain period.
An example of failures that cause the level-shift pattern is packet drops.
This is because packet drops partially reduce traffic volume.
The third pattern is ramp down in Figure~\ref{fig:traffic_pattern} (c), which gradually reduces traffic volume.
An example of the failures that cause ramp down pattern is deterioration of module in the network element.
This is because deterioration of the module gradually progresses and the traffic in the deteriorated network element gradually reduces.
%Moreover this deterioration is more difficult for network operators to recognize than the failures above.
In addition to the degradation patterns introduced by Mahimkar et al.~\cite{Mahimkar2011a}, we added another degradation pattern of the traffic volumes.
The fourth pattern is {\it long-period down} in Figure~\ref{fig:traffic_pattern} (d), which completely reduces traffic volume.
An example of the failures that cause long-period down is a hardware failure.
This is because a hardware failure completely stop sending or receiving traffic.

\subsection{Synthetic Traffic Volumes and Syslog}
In this subsection, we explain how traffic volume and syslog messages in normal status were generated and then explain how degradation were injected into the data in normal status.

\subsubsection{Synthetic Data in Normal Status}
We created synthetic traffic volumes by superimposing sine waves with several frequencies.
Typically, traffic volumes increase early in the morning, decrease in the afternoon, and then increase in the evening and around night again.
This is because office workers use the network when commuting in the morning and evening.
Many people also use the network during their leisure time such as for watching videos after work.
Therefore, we superimposed sine waves so that a superimposed wave has peak traffic in the evening and night.
%We also superimposed sine waves with short frequencies of 5 min., 10 min., 30 min., 2 hour and 5 hour to make traffic volumes,
%since traffic volume also depends on sine waves with short frequencies.
%This is normal behavior of the traffic volume for fundamental experiments.
We also added white Gaussian noise to the superimposed wave.

We next explain how to generate synthetic syslog data.
For synthetic syslog data, we did not generate syslog messages one by one,
but generated vectors $\{\boldsymbol{x}_t\}^T_{t=1}$ described in Section~\ref{sec:proposal_method}.
%The $i$-th element of each vector $x_t$ represents the number of occurrences of syslog messages to which template ID $i$ is assigned.
%We generated these vectors with the settings described below.
%The $i$-th element of the vector represents the number of occurrence of syslog message to which template ID $i$ is assigned.
%A sample of prepared vector to imitate syslog messages in the network element is illustrated in Figure~\ref{fig:prepared_vector}
To generate syslog in normal status, we categorize typical generation patterns of syslog into {\it periodic} and {\it random} events,
and generated vectors on the  basis of each generation pattern.
A syslog message is periodically generated when a periodic event occurs such as a backup system running.
Syslog messages are also randomly generated when the user event occurs such as a user log-in.
%Moreover, generated syslog messages of periodic patterns are different from those of random pattern.
%For instance, if a backup system runs every 2 min, a certain syslog message will be generated every 2 min.
%Similarly, when we log in to a network element, messages related to log-in will be generated.
To imitate periodic syslog generation, we prepared 10 template IDs for syslog which are generated by periodic events
with every 2 min, 3 min, 5 min, 10 min, 30 min, 1 hour, 2 hours, 5 hours, 12 hours, and 24 hours, respectively.
To imitate random syslog generation, we generated 30 other template IDs for syslog that are generated by random events.
In this way, we generated vectors in normal status to imitate syslog messages in the network element.
The generated syslog vectors correspond to approximately 1000 messages per day.
%\begin{figure}
%\centering
%\includegraphics[width=0.4\textwidth]{figs/syntheticlogsample.pdf}
%\caption{A sample of prepared vector to imitate syslog}
%\label{fig:prepared_vector}
%\end{figure}

\subsubsection{Synthetic Data in Abnormal Status}
We next explain how to generate synthetic traffic volumes and syslog messages in abnormal status.
First, we generated synthetic traffic volume at the time a failure occurred by injecting the degradation patterns into synthetic traffic volumes in normal status and set recovery time on the basis of recovery procedures described in Subsection~\ref{subsec:degradation_patterns}.
%We introduced four patterns of degradation of traffic volumes and corresponding failures,
%which are spike (link flap), level-shift (packet drops), ramp down (module deterioration), and long-period down (hardware failure) based on~\cite{Mahimkar2011a}.
%To generate traffic volumes in abnormal status, we reduced traffic volumes of normal status and set recovery time
%on the basis of the degradation patterns and recovery procedures described in SubSection~\ref{subsec:degradation_patterns}.
\begin{table*}[t!]
\centering
\caption{Summary of synthetic dataset}
\label{tab:dataset}
  \begin{tabular}{c|c|c|c|c}
  \hline
  Traffic behavior & Example of failure & Time To Recovery & Traffic reduction & Number \\
  \hline\hline
  Spike & Interface down  & 5min & Complete reduction & 41-60 \\
  \hline
  Level-shift & Packet loss & 10min & 50\% Reduction & 61-80 \\
  \hline
  Ramp down & Module deterioration & 60min & 1\% Reduction per 1 min. & 1-40 \\
  \hline
  Long-periodic down & Hardware failure & 120min & Complete reduction & 81-100\\
  \hline
  \end{tabular}
\vspace{5mm}
\end{table*}
Then, we prepared syslog when a failure occurred by adding other template IDs to syslog vectors in normal status.
When a failure occurs, syslog messages that are different from syslog messages in the normal status are typically generated.
These syslog messages for each failure are also different from each other except those for ramp down.
Therefore, we prepared 20 template IDs for each failure that represent syslog messages at time $T$.
For ramp down, we randomly change generation patterns of periodic and random syslog messages.
Although failures such as deterioration of the module do not completely stop systems, these failures make difficult for network operators to run back-up systems or log in to the system.
Therefore, syslog messages might not be generated, but the frequencies of periodic syslog messages or random syslog messages in the network elements might be changed due to the failure.

The parameters of abnormal traffic patterns and syslog messages are summarized in~Table~\ref{tab:dataset}.
We prepared synthetic traffic volumes and vectors in normal status for 6 days, and randomly injected abnormal behavior during the 6th day.
The duration of a time slot is set as 1 min,
With these settings, we prepared 9000 data for training and 1000 data for evaluation.

Since DeepSIP needs a number of data for training, we used the synthetic data for evaluation.
However, we can also prepare a training dataset by injected the degradation patterns to real traffic data, as in the paper~\cite{Mahimkar2011a}.
%, we can generate training dataset by injected the degradation patterns to real traffic data. 
We can also apply DeepSIP trained in a test network environment to a real network environment by using a transfer learning approach~\cite{weiss2016survey}.
A data augmentation technique~\cite{shorten2019survey} also enables us to increase the number of training dataset obtained from a real network. 
Therefore, proposed system is feasible and enables to apply to a real network operations.

\subsection{Evaluation Results}
\begin{figure}
\centering
\includegraphics[width=0.4\textwidth]{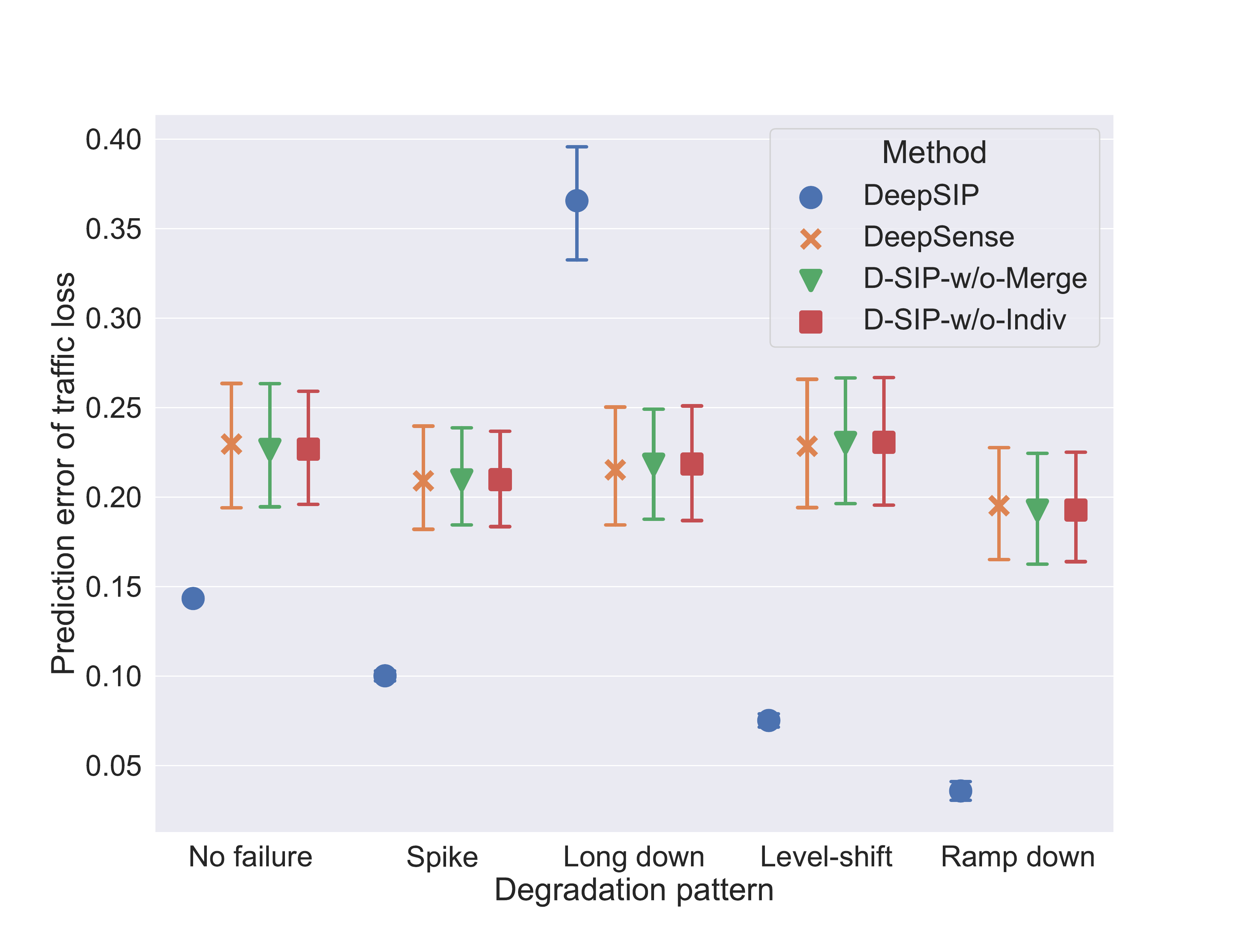}
\caption{Prediction error of $V_{\mathsf{TTR}}$}
\label{fig:error_of_traffic_loss}
\end{figure}
\begin{figure}
\centering
\includegraphics[width=0.4\textwidth]{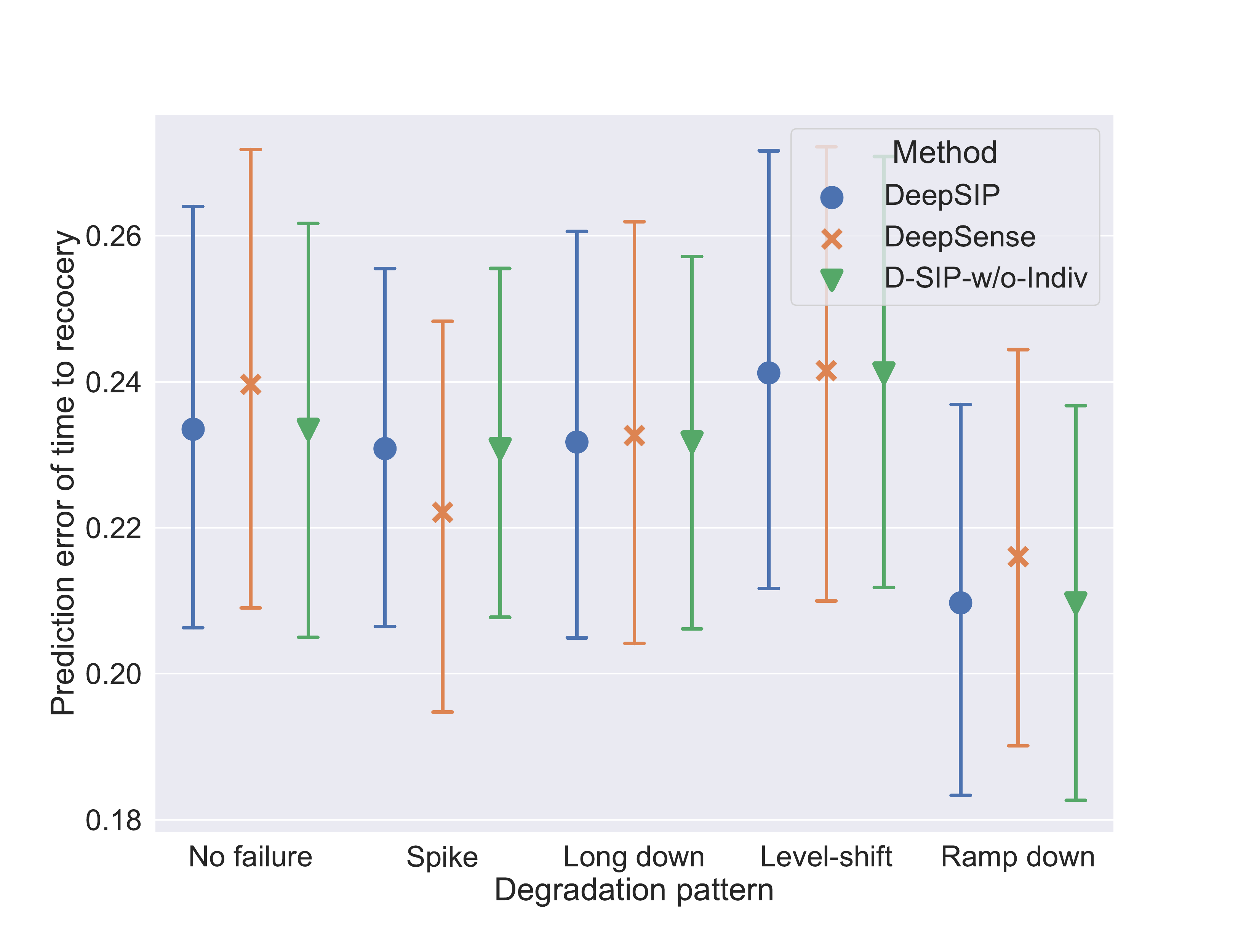}
\caption{Prediction error of $\mathsf{TTR}$}
\label{fig:error_of_time_to_recovery}
\end{figure}

Using the synthetic dataset, we evaluated the accuracy and effectiveness of DeepSIP.
We implemented DeepSIP by Pytorch and used NVIDIA Tesla 100 GPU for training and evaluation.
%The computation time was roughly 2 hours for training with 100 epochs, but this is an off-line process.
%The prediction phase needs less than 1 min.

We considered three different architectures as base-line: DeepSense, DeepSIP without merge CNN (D-SIP-w/o-Merge), DeepSIP without individual CNN (D-SIP-w/o-Indiv).
Figures~\ref{fig:error_of_traffic_loss}  and~\ref{fig:error_of_time_to_recovery} show mean square errors and 95\% confidence intervals.
%They show the error of prediction for time to recovery and loss of traffic volume in DeepSIPs and DeepSence.
We set the prediction error as the relative error between the prediction and the ground truth.
Figure~\ref{fig:error_of_traffic_loss} illustrates prediction results of $V_{\mathsf{TTR}}$.
For prediction of $V_{\mathsf{TTR}}$, the results of DeepSIP were better than those of base-lines for every degradation pattern except long down pattern.
For each pattern except long down pattern, DeepSIP reduced the prediction error by approximately 50\% and the deviations of prediction errors of DeepSIP were quite smaller than those of other methods.
This means that extracting the features using individual and merge CNN improved the prediction.
%Extracting the features using individual and merge CNN improved the prediction.
DeepSIP reduced the prediction error comparing with DeepSense.
This means that QRNN in DeepSIP also extracted temporal dependencies better than the GRU in DeepSense.
Since the prediction errors of DeepSIP were quite small, this information is useful for network operators to prioritize failures and handle SLA.

Figure~\ref{fig:error_of_time_to_recovery} illustrates prediction results of $TTR$.
Since D-SIP-w/o-Merge did not converge to this problem, we eliminated the results of D-SIP-w/o-Merge from this Figure.
This means that extracting the features using merge CNN improved the prediction.
For prediction error of $\mathsf{TTR}$, the results of DeepSIP were similar to the those of base-lines for every degradation patterns.
For spike, long down and ramp down pattern, the results of DeepSIP were slightly better than those of base-lines.
This means that individual CNN and QRNN in DeepSIP improved the predictions.
The prediction errors of DeepSIP are 10 seconds for all degradation patterns.
Thus this information is useful for network operators to prioritize failures and handle SLA.

\section{Conclusion}
\label{sec:conclusino}
Predicting the service impact is important for network operators since service impact is useful information for SLA and determining recovery procedure.
%Therefore, in this paper, we proposed DeepSIP, a system to automatically predicts service impact using syslog messages and traffic volume and temporal multimodal CNN.
In this paper, to analyze syslog messages and traffic volume which contain useful information for predicting the time to recovery and the loss of traffic volume,
we used temporal and multimodal CNN.
%We evaluated DeepSIP by fundamental and realistic synthetic data, which imitated the traffic volume and syslog messages.
%Prediction of DeepSIP was best for both dataset.
In the experiments, DeepSIP reduced prediction error by approximately 50\% in comparison with base-lines.
We showed the effectiveness of temporal multimodal CNN in DeepSIP by comparing with DeepSIP without individual CNN or merge CNN and DeepSense.
Future works includes improving the prediction by customizing the loss function of DeepSIP and evaluating the effectiveness of DeepSIP with a real dataset.
%The following remains for future work.
%First, we will use more data collected from network elements such as show tech logs for extracting hidden information about failures.
%As we described, syslog messages contain the hidden information about failures.
%However, it does not includes telemetry data such as fan speed or temperature of module in network element.
%Thus, telemetry data is useful for predicting service impact.
%Second is improving the prediction by customizing the loss function of DeepSIP.
%We use mean squared error as loss function.
%Second is more evaluation using datasets in large network.
%Although we comprehensively prepared synthetic dataset for evaluation,
%we should validate effectiveness of DeepSIP with real dataset.

\end{document}